\documentclass[usenatbib]{mn2e}
\usepackage[dvips]{color}
\usepackage{graphicx}
\usepackage{lscape}

\def\*{$^{*}$}
\def\ergs{erg~s$^{-1}$}

\title[Strong outburst activity of X\,Persei]
{Strong outburst activity of the X-ray pulsar X\,Persei during 2001-2011}
\author[Lutovinov, Tsygankov, Chernyakova]{A.\,Lutovinov$^{1}$\thanks{E-mail:
aal@iki.rssi.ru}, S.\,Tsygankov$^{2,3,1}$, M.\,Chernyakova$^{4,5}$\\
$^{1}$Space Research Institute of the Russian Academy of Sciences, Profsoyuznaya Str. 84/32, Moscow
  117997, Russia\\
$^{2}$Finnish Centre for Astronomy with ESO (FINCA), University of Turku,  V\"ais\"al\"antie 20, FI-21500 Piikki\"o, Finland \\
$^{3}$Astronomy Division, Department of Physics, FI-90014 University of Oulu, Finland\\
$^{4}$Dublin City University, Dublin 9, Ireland\\
$^{5}$Dublin Institute for Advanced Studies, 31 Fitzwilliam Place, Dublin 2, Ireland}

\begin{document}

\date{Accepted .... Received ...}

\pagerange{\pageref{firstpage}--\pageref{lastpage}} \pubyear{2012}

\maketitle

\label{firstpage}

\begin{abstract}
{We present results of a comprehensive analysis of the X-ray pulsar X\,Persei over the period 1996 to 2011, encompassing the quite low state and subsequent strong outburst activity. Using data from the \textit{RXTE} and \textit{Swift} observatories we detected several consecutive outbursts, during which the source luminosity increased by a factor of $\sim5$ up to $L_X\simeq1.2\times10^{35}$ \ergs. Previously the source was observed in a high state only once. The source spectrum in a standard energy band ($4-25$ keV) is independent of the flux change and can be described by a model that includes both thermal and non-thermal components. With the help of the \textit{INTEGRAL} observatory data we registered the highly significant cyclotron absorption line in the source spectrum and, for the first time, significantly detected a hard X-ray emission from the pulsar up to $\sim160$ keV. We also report drastic changes of the pulse period during the outburst activity: a long episode of  spin-down was changed to  spin-up with a rate of $\dot P/P \simeq -(3-5)\times10^{-4}$ yr$^{-1}$, that is several times higher than previous rates of spin-up and spin-down. To search for a correlation between the X-ray and optical lightcurves we took data from the AAVSO International Database. No significant correlation between optical and X-ray fluxes at any time lag from dozens of days to years was found. }

\end{abstract}


\section{Introduction}

4U\,0352+309/X\,Persei is a classical persistent Be/X-ray binary system, consisting of an X-ray pulsar and a
Be star companion optically identified with the star HD\,24534. It was discovered during a high X-ray intensity
state in 1972 (for simplicity below  such high states are referred to as outbursts), when pulsations with a
period of $\sim835$ s were detected by the \textit{Copernicus} observatory \citep{wht76}. The distance to the source
was estimated by different authors in the range of 700 to 1300 pc. In this paper we used the value of
$950\pm200$ pc obtained by \citet{tel98}. Adopting this distance the source peak luminosity $L_X\simeq2\times10^{35}$ \ergs\ in the $2-10$ keV energy range was registered in 1975. The close proximity  allowed the source to be observed in   X-rays by different missions and experiments after 1978, despite its relative low luminosity $L_X\simeq(2-4)\times10^{34}$ \ergs.

\bigskip

For a long time X\,Persei was the only known persistent X-ray pulsar in Be-systems, until
\cite{reig1999} pointed out the existence of a subclass of binaries with Be
companions which harbor a slowly rotating neutron star and are characterized
by a persistent low-luminosity X-ray emission and, probably, a long orbital period.
Soon after, \citet{delma01} succeeded in determining orbital parameters for X\,Persei and showed
that the pulsar is in a moderately eccentric orbit ($e=0.11$) with a $P_{orb}\simeq250$ days orbital period.
Recently, \citet{tsy11} reported a similarly long orbital period of $\simeq155$ days for another
X-ray pulsar RX\,J0440.9+4431, which belongs to this subclass as well.

Usually spectra of X-ray pulsars have an exponential cutoff at energies about $\sim$20 keV
\citep[see, e.g., the recent review of][and references therein]{fil2005}. However results of the \textit{Ginga}
observatory gave evidence for the presence of a hard tail in the spectrum of X\,Persei  above $\sim15$ keV\citep{rob96}.
Based on data of the \textit{BeppoSAX} observatory, \citet{disal98} showed that its broadband spectrum
can be described by a combination of the standard model and a power law with both high- and low-energy cut-offs,
which dominates above $\sim 15$ keV. Later \citet{lapal07} suggested that the thermal component
with temperature of $kT>1$ keV is typical for spectra of low luminosity, long period Be/NS binaries.

Similar to other accreting pulsars, the history of the X\,Persei pulse period
shows episodes of both spin-up and spin-down. Previous to 1978 its spin
history was erratic, with a general spin-up trend at a rate of $\dot
P/P\simeq-1.5\times10^{-4}$ yr$^{-1}$. Since 1978 the neutron star has been
spinning down with a similar rate $\dot P/P\simeq1.3\times10^{-4}$ yr$^{-1}$
\citep{delma01}. Note that the spin-up episode in 1975-1978 coincided with the
decay phase of the outburst.

Be stars are known to have a dense slow disk-like equatorial wind along with a low-density fast wind at
higher latitudes. X\,Persei has shown large variations in optical and infrared brightness ($V = 6.1-6.8$,
$K = 5.2-6.7$) and in the strength of the emission Balmer line \citep[see, e.g.][and references therein]{roch93,tel98,cla01}.
These variations are usually interpreted as changes in the physical properties (density, geometry) of the Be star equatorial
disk . The bright state of the star correlates with the presence
of the emission lines and is evidence of the equatorial disk presence. At the same time, as it was mentioned
by \citet{tel98}, the strong variability seen in the optical and infrared is not reflected in the X-ray light curve.
Particularly, the 1975 peak of the X-ray flux happened during the diskless state, while during the 1990 diskless
state no X-ray brightening was observed. The absence of a direct correlation between the optical and X-ray light
curves excludes the possibility that the neutron star is always accreting from the Be disc, probably due to
the truncation of the disc \citep{cla01}. On the other hand the detailed study of the correlation between
optical and X-rays was hampered by the absence of continuous measurements of the source flux in X-rays.

In this paper we present a comprehensive analysis of the properties of the X-ray pulsar X\,Persei over the period 1996 to 2011, encompassing the quite low state and subsequent strong outbursts activity. Thanks to the \textit{RXTE} and
\textit{Swift} observatories, which have regularly monitored the whole sky in the last decades,
we are able to look for the first time for a correlation between X-ray and optical data on a time scale from days to years.
Moreover, based on data from the \textit{INTEGRAL} and \textit{RXTE} observatories we traced the evolution of the source pulse period and its spectral parameters during outbursts. We also used deep observations of the \textit{INTEGRAL} observatory to detect for the first time a hard X-ray emission from the pulsar up to $\sim160$ keV at a high significance level.

\section{Observations and Data Analysis}

In this work we use data obtained with the \textit{RXTE} \citep{br93} and \textit{INTEGRAL} \citep{win03} observatories. In particular, we used data of the \textit{ASM/RXTE} monitor to trace the long-term behavior of the X\,Persei flux in the soft energy band ($<12$ keV). To compare it with the source behavior in hard X-rays we used observations of the \textit{IBIS/INTEGRAL} and \textit{BAT/Swift} telescopes \citep{gehr2004}. In the standard X-ray energy band ($4-25$ keV) data from the \textit{PCA/RXTE} instrument (Obs.\,ID. 30099, 40424, 50404, 60067, 60068) were used to study the variability of the source spectrum and pulse period.

In 2003-2006 X\,Persei was in the field of view of telescopes of the INTEGRAL observatory only episodically.
During deep observations of sky fields around the source in 2008-2010 the effective exposure of both X-ray
telescopes of the observatory, \textit{JEM-X} and \textit{IBIS/ISGRI}, rose up to $\simeq147$ ksec and $\simeq3.7$ Msec, respectively. The reduction of the \textit{ISGRI} detector data was done using recently developed methods described by \cite{kri2010}. The source spectrum from the \textit{JEM-X} telescope was extracted with the standard \textit{OSA} package version 9.0\footnote{http://isdc.unige.ch}. For timing analysis the \textit{IBIS} data was processed  with the software developed and maintained in the National Astrophysical Institute in Palermo\footnote{http://www.pa.iasf.cnr.it/~ferrigno/INTEGRALsoftware.html)}; a description of the data processing technique can be found in \citet{min06} and \citet{seg07}.

To build a complete picture of the secular variability of the X\,Persei pulse period we included in our analysis data from the \textit{XMM-Newton} observatory that observed the source twice in Feb 2003 and Feb 2010 (Obs.\,ID. 0151380101 and 0600980101, respectively). Version 11.0 of the \emph{XMM-Newton Science Analysis System SAS} was used to process the data.

The final spectral and timing analysis for all X-ray instruments was done using the standard tools of the
FTOOLS/LHEASOFT 6.7 package.

To search for a correlation between the source outbursts activity in X-rays and optics ($V$-band), we took data from the AAVSO International Database\footnote{http://www.aavso.org}.

\section{Results}

\subsection{X-ray light curves}

\begin{figure}
\includegraphics[width=\columnwidth,bb=20 140 560 710,clip]{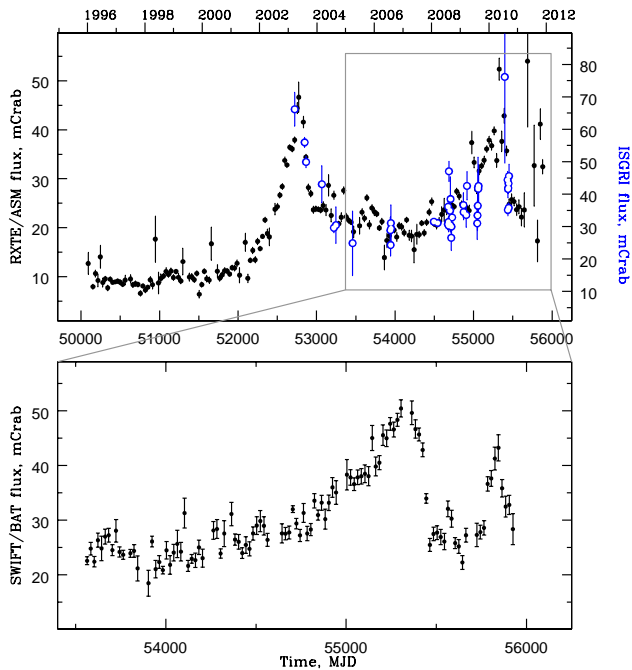}

\caption{({\it top}) Light curves of X\,Persei obtained with the \textit{ASM/RXTE} monitor in $1.3-12$ keV (filled circles) and \textit{IBIS/INTEGRAL} telescope in $20-60$ keV (open circles) energy bands. The grey box indicates an interval with simultaneous \textit{BAT/Swift} observations. ({\it bottom}) Corresponding light curve of the source obtained with the \textit{BAT/Swift} telescope in the 15-50 keV energy band. To optimize the signal-to-noise ratio light curves were averaged over 30 and 20 days for \textit{ASM/RXTE} and \textit{BAT/Swift}, respectively, while \textit{ISGRI/INTEGRAL} measurements were averaged over one spacecraft revolution (about three days). \label{fig:xray_lc}}
\end{figure}

The X\,Persei light curve obtained with the \textit{ASM/RXTE} monitor in 1996-2011 indicates that during the first years of observations the source intensity was low and relatively stable at the level of $\simeq10$ mCrab, that corresponds to a luminosity of $L_X\simeq2.3\times10^{34}$ erg s$^{-1}$ in the $2-10$ keV energy band for the adopted distance (Fig.\ref{fig:xray_lc}). Such a level of intensity is typical for the normal (low) state of the pulsar, which had been observed since the end of 1970s \citep[see, e.g.][and references therein]{roch93,tel98}. In 2000-2001 the source flux began to steadily increase and reached a maximum ($\simeq50$ mCrab) in May 2003. It then dropped very quickly (within 4-5 months) by a factor of two and stayed near this level until 2008, when it began to increase again. The source flux at the new maximum (May 2010) was approximately the same as the preceding one. And again it dropped very quickly by a factor of 2 (Fig.\ref{fig:xray_lc}).

The described source behavior at soft energies , as registered by the \textit{INTEGRAL} (20-60 keV) and \textit{Swift} (15-50 keV) observatories (Fig.\ref{fig:xray_lc}), almost completely mirrors the source variability  in hard X-rays. The source outburst activity continued during 2011, when a third outburst with a slightly lower intensity was detected by the \textit{Swift} observatory. A comparison of X-ray light curves of X\,Persei from
Fig.\ref{fig:xray_lc} with the source long-term behavior reported by \citet{roch93,tel98} shows that a similar
outburst activity was observed more than 35 years ago, in the middle of 1970s, with the \textit{Ariel V} and \textit{Copernicus} observatories. Moreover, the maximal flux from the source in 1975, was even slightly higher that  measured in 2003 and 2010 and corresponded to a source luminosity of $L_X\sim2\times10^{35}$ erg s$^{-1}$ \citep{tel98}.

Besides these large outbursts with durations of years, the light curves of X\,Persei demonstrate a tentative variability with the amplitude of about $5-10$ mCrab. To search for a possible correlation of these source intensity variations with the orbital phase, we folded \textit{ASM/RXTE} and \textit{BAT/Swift} light curves with a period of $P_{orb}=250.3$ days \citep{delma01}, but we did not find any significant deviations from the constant either in soft or in hard X-rays. Finally, note that the intensity rise, which has been very smoothly for several hundred days for all observed outbursts from X\,Persei,  is non-typical for outbursts from usual transient sources (like black holes or transient X-ray pulsars). It can be connected with other mechanisms or processes responsible for the outburst activity in the source under investigation.

\subsection{Pulse period evolution}
\begin{figure}
\includegraphics[width=\columnwidth,bb=105 135 565 675,clip]{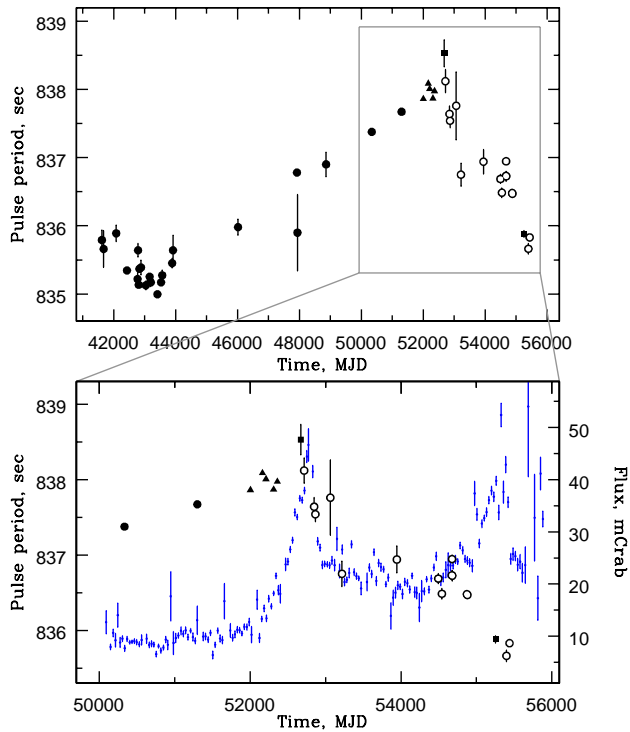}

\caption{Secular variability of the X\,Persei pulse period as historically observed by
\citet{nag89,lut94,hab94,rob96,disal98,delma01} (dark circles). Results of \textit{RXTE}, \textit{INTEGRAL} and
\textit{XMM-Newton} measurements from this paper are shown by the triangles, open circles and dark squares,
respectively. \label{fig:pulseper}}

\end{figure}

The secular variability of the X\,Persei pulse period is presented in Fig.\ref{fig:pulseper}. It includes results of measurements taken from the literature and calculations performed in this work. The photon arrival times have been preliminarily corrected to the barycenter of the Solar System. After that the epoch folding method has been applied to determine the period of the pulsar. The errors in the period were estimated by the method briefly described by \citet{fil04}. To calculate them we performed an analysis of $10^3$ simulated light curves. Data in each light curve were generated arbitrarily inside the error bars of the original data points. After that the best period for each simulated light curve was determined. Finally we computed the standard deviation of the best periods distribution and used this value as an estimation of the pulse period error.

Due to the  faintness of the source and a relatively long pulse period ($\sim14$ min) observations with an exposure of several hours (at least) are needed to determine the pulse period somewhat accurately. This was not a problem for \textit{INTEGRAL} and \textit{XMM-Newton} observations, where exposures were about several dozen ksec, but the typical duration of one \textit{RXTE} observation is only several ksec. Therefore we grouped closely spaced  \textit{RXTE} observations into one set (when possible) and calculated the pulse period for each of them. Data of the \textit{XMM-Newton} observatory were highly affected by event pile-up due to a very high count rate. Therefore in our analysis we used the same approach as \citet{lapal07} and considered only data from the annulus area, excluding the central part of the source PSF. Applying the standard processing described in the \emph{SAS} v.11 cookbook we determined the pulse period of X\,Persei: $838.52\pm0.20$ and $835.94\pm0.05$ sec for February 2003 and February 2010, respectively. The quality of the data during the last observation was much better than during the first one as its duration was longer ($\sim31$ ksec vs $\sim126$ ksec), therefore the corresponding uncertainty is lower.  The obtained value of the pulse period in February 2003
is slightly lower than the value reported by \citet{lapal07}. There are two possible reasons for this difference: we used a more recent version of \emph{SAS}, and we used the data from \emph{MOS} cameras only. As mentioned by \citet{lapal07} data of the $pn$ camera were affected not only by  photon pile-up, but during that observation the source was imaged close to a CCD gap. Finally, note that our value of the X\,Persei pulse period for February 2003 agrees well with the measurements from the \textit{INTEGRAL} observatory, performed in April 2003 (see Fig.\ref{fig:pulseper}).

Measurements from the \textit{INTEGRAL} and \textit{XMM-Newton} observatories indicate that sometime before the 2003 source flux maximum a deceleration of the neutron star rotation, which had lasted from 1978, changed to an acceleration (Fig.\ref{fig:pulseper}). This repeats approximately the pulse period behavior during the outburst in 1974-1978, when the spin-up episode coincided with its declining phase, but started $\sim400-500$ days after the luminosity maximum \citep{roch93}. The average spin-up rate since 2003 $\dot P/P \simeq -3.6\times10^{-4}$ yr$^{-1}$ is higher than the one observed between 1974-1978. Moreover, measurements of the \textit{INTEGRAL} observatory suggest the presence of something like a plateau in the pulse period history during the time interval MJD $53000-54500$, when the source flux was more or less stable between the first and second outbursts. In such a case the pulsar is spinning-up only during outbursts with the even higher rate of $\dot P/P \simeq -5\times10^{-4}$ yr$^{-1}$. The high spin-up rate can be probably connected with additional matter and angular momentum transfer to the neutron star due to the continuing  source high state after the first outburst, and subsequent outbursts. This picture can be naturally explained in terms of the current theory of quasi-spherical accretion in low-luminosity X-ray pulsars \citep{sha12}. In particular, the observed spin-up/spin-down transitions correlate with changes of the source luminosity, and their numerical values reflect the ratio of X-ray fluxes during spin-up and spin-down \citep[see Fig.1 and eqs. 56-58 in][]{sha12}. Moreover, the plateau at the pulse period history is apparently connected with the equilibrium state of the pulsar. Using the equation $$P_{eq}\approx10^3 \mu_{30}^{12/11} \dot M^{-4/11}_{16} \left(\frac{P_{orb}}{10d}\right) v^{4}_{8} ~~\rm{s},$$ where $P_{eq}\simeq837$ s is the equilibrium period, $\mu_{30}$ is the neutron star magnetic moment in units $10^{30}$ G cm$^3$, which can be calculated from the measured magnetic field (see below), $\dot M_{16}$ is the accretion rate in units $10^{16}$ g/s, deriving from the source luminosity, and $v_{8}$ is the relative wind velocity (in units $10^8$ cm/c), we can estimate  $v_{8}$ as $\sim200$ km/s, which agrees with  measured or suggested values \citep[see, e.g.,][]{wat88,delma01}.

As noted by \citet{lut04}, pulsations from the source are detected up to 100 keV. The pulse fraction
increases from $\sim30$\% in the $2-10$ keV energy band to $\sim60$\% in the $20-60$ keV energy band, typical for X-ray pulsars \citep{lut09}.

\subsection{Evolution of the spectral parameters}
\begin{figure}
\includegraphics[width=\columnwidth,bb=50 165 550 690,clip]{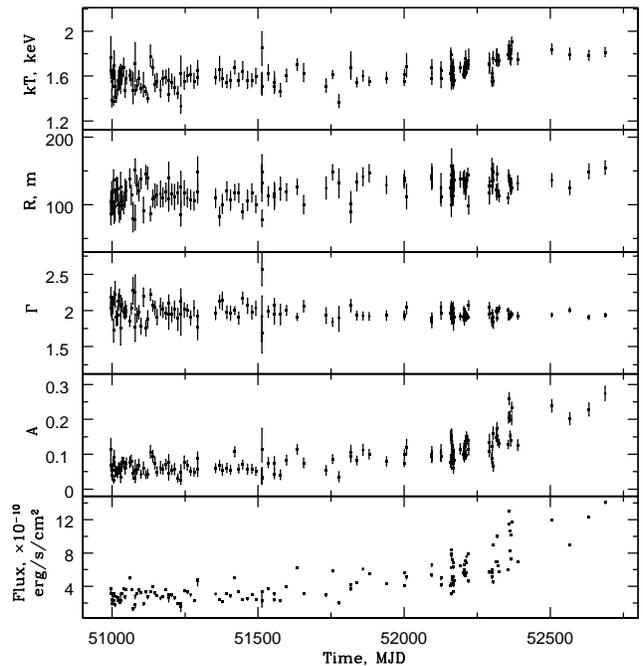}

\caption{Variability of spectral parameters and the $2-20$ keV model flux of X\,Persei during the rising part of the first outburst as measured with the \textit{PCA/RXTE} instrument (see text for details).
\label{fig:pca_par1}}

\end{figure}

Observations of X\,Persei with the \textit{PCA} instrument of the \textit{RXTE} observatory were performed several
dozens of times in 2000-2003. They  cover the low state and the rising phase of the first outburst very well. It allowed us
to study a dependence of the source spectral shape  variability on its intensity.

Using \textit{RXTE/PCA} data we built a set of source spectra in the $4-25$ keV energy band and approximated them
in a model that includes  interstellar absorption, powerlaw and black-body components (\emph{wabs, powerlaw and bbodyrad} models in the \textit{XSPEC} package). Interstellar absorption in the direction of the source is quite low ($N_{\rm H}=few\times10^{21}$) cm$^{-2}$ and cannot be determined correctly with the \textit{PCA} instrument. It was fixed at the value of $N_{\rm H}=3.4\times10^{21}$ cm$^{-2}$ as measured by
\textit{XMM-Newton} \citep{lapal07}. Dependencies of the black-body temperature $kT$, the radius of the emission surface of the thermal component $R$, the photon index $\Gamma$ and the normalization of the non-thermal component $A$ with time are shown in Fig.\ref{fig:pca_par1} along with the changes of the source flux in the $2-20$ keV energy band, which was calculated from the best fit spectral models.

It is clearly seen that the photon index remains very stable during the whole period of observations and its value $\Gamma\simeq2$ is in a good agreement with broadband spectral measurements performed with the \textit{INTEGRAL} observatory (see below). At the same time, the normalization of the non-thermal component $A$ increases with the source brightness by a factor of $3-4$. This increase is more significant in comparison with the extension of the emission surface of the thermal component from $R\simeq110$ m to $R\simeq150$ m that should lead to the increase in the relative contribution of the non-thermal component to the total source spectrum with the source brightening. But quantitatively this increase is practically negligible due to the simultaneous growth of the black-body temperature $kT$ from $\sim1.6$ to $\sim1.8$ keV (Fig.\ref{fig:pca_par1}), that in turn leads to the growth of the thermal component flux.

It is important to note that measured values of the thermal component parameters (black-body temperature and radius) agree with previous results of \citet{cob01}, but are different from the ones obtained with \textit{XMM-Newton} \citep{lapal07} -- the temperature is slightly higher, the radius is lower by a factor of two. This difference is due to the different energy bands of the observatories: $0.3-10$ keV for \textit{XMM-Newton} and $4-25$ keV for \textit{RXTE}. Naturally, the capabilities of the \textit{XMM-Newton} observatory are much more suitable for a correct determination of quantitative characteristics of the thermal emission with temperatures of $kT\sim(1-2)$ keV. Therefore the \textit{RXTE} data reflect only the qualitative behavior (relative changes) of its parameters as the source flux increases.

In addition to the long-term changes, the source X-ray flux demonstrates variability on the time scale of several days, which became especially significant (like flares) with the growth of the source average flux \citep[Fig.\ref{fig:pca_par1}, bottom panel, see also Fig.2 from][]{delma01}. Such flares are accompanied with the variability of the source spectral parameters, mainly its thermal component. Note that flares with changes of the spectrum hardness were observed earlier during the high luminosity states in 1970s \citep{wht76,fro79}.

\subsection{Hard X-ray emission and cyclotron line}

\begin{figure}
\includegraphics[width=\columnwidth,bb=50 235 550 710,clip]{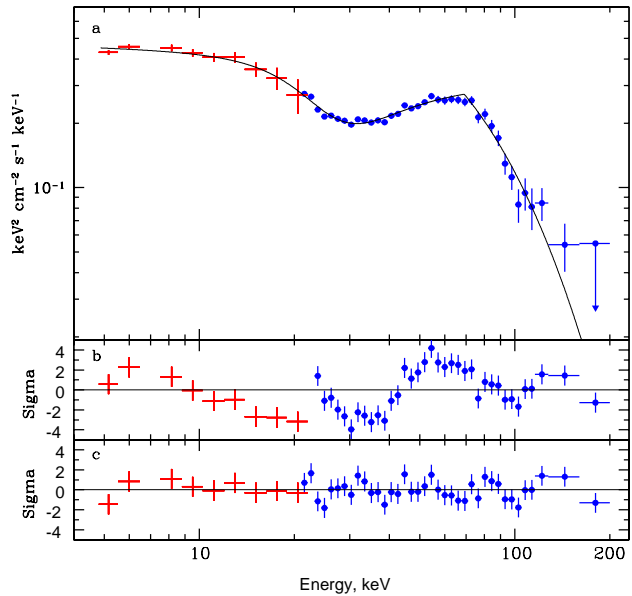}

\caption{Broadband energy spectrum of X\,Persei obtained with the \textit{JEM-X} (crosses) and \textit{IBIS} (dark circles) telescopes of the \textit{INTEGRAL} observatory. Solid line represents the best-fit model with the cyclotron absorption line in the form of \textit{cyclabs} ({\it a}). The residuals to the fit in units of $\sigma$ without ({\it b}) and with ({\it c}) a cyclotron absorption line (see text for details). \label{fig:int_spec}} \end{figure}

X\,Persei was observed by the \textit{INTEGRAL} observatory with a very long exposure in 2008-2010. Taking into account that 1) the source intensity didn't change significantly during these observations (Fig.\ref{fig:xray_lc}) and that 2) the hard non-thermal spectral component is very stable even when the source flux is variable (see previous section), we could use all these data to reconstruct the broadband spectrum of X\,Persei. In the resulting spectrum,   the signal from the source is significantly detected up to $160$ keV
(Fig.\ref{fig:int_spec}a) for the first time. To date, hard X-ray emission at energies $>100$ keV has been detected from only one other low luminosity X-ray pulsar RX\,J0440.9+4431 \citep{tsy12}.

As a first step the source spectrum was approximated by the powerlaw model with a high-energy cutoff that is
typical for X-ray pulsars \citep{wht83}\footnote{There is no necessity to include into the model an
interstellar absorption and a thermal component with the temperature of $kT\sim1.5$ keV as the sensitivity of
the \textit{JEM-X} telescope is much lower in a comparison with \textit{PCA/RXTE}. Moreover,
due to problems with the \textit{JEM-X} response matrix energy channels around 6-7 keV were
excluded from the analysis according to recommendations of the ISDC helpdesk (private communication)}.
However this model gives an unacceptable value of $\chi^2=275$ for 45 d.o.f. Deviations of the measured spectrum from the model (Fig.\ref{fig:int_spec}b)
demonstrate a wide absorption feature near $30$\,keV, which can be interpreted as a cyclotron resonance
absorption line and which was found earlier by \citet{cob01} based on the \textit{RXTE} data.

An additional component in the form of \emph{gabs} (\textit{XSPEC}):
$\exp\left(\left(\frac{-\tau_\mathrm{cycl}}{\sqrt{2\pi}\sigma_\mathrm{cycl}}\right)
\exp\left(\frac{-(E-E_\mathrm{cycl})^2}{2\sigma_\mathrm{cycl}^2}\right)\right)$,
improves the fit significantly to $\chi^2=58.2$ for 42 d.o.f. The resulting parameters of the cyclotron ine -- the line energy $E_{cycl}=29.0\pm1.0$ keV and the line width $\sigma=9.9\pm1.0$ keV -- are in a good agreement with the results of \citet{cob01}, who also used the \emph{gabs} model. Using another suitable model \emph{cyclabs} (\textit{XSPEC}): $\exp\left(\frac{-\tau_\mathrm{cycl}(E/E_\mathrm{cycl})^2\sigma_\mathrm{cycl}^2}
{(E-E_\mathrm{cycl})^2+\sigma_\mathrm{cycl}^2}\right)$ for the description of the cyclotron absorption line improves the fit even more to $\chi^2=43.7$ for 42 d.o.f. (Fig.\ref{fig:int_spec}c); however the line energy is slightly lower in this fit ($\sim23.5$ keV), typical for a cyclotron absorption line that is approximated by \emph{gabs} or \emph{cyclabs} models.

Thus, the parameters of the best-fit model (with the \emph{cyclabs} component), describing the broadband spectrum of X\,Persei in the $4-200$ keV energy range are:  photon index $\Gamma=2.05^{+0.05}_{-0.04}$, cutoff energy $E_{cut}=69^{+2}_{-3}$ keV, folding energy $E_{fold}=34.0^{+3.0}_{-2.5}$ keV,  line energy $E_{cycl}=23.5^{+1.5}_{-1.4}$ keV,  line width $\sigma=13.4^{+1.1}_{-1.4}$ keV,  line depth $\tau=0.56^{+0.04}_{-0.05}$. Based on the measured value of the cyclotron line energy we can estimate the magnetic field on the neutron star surface as

$$B_{NS}=\frac{1}{\sqrt{\left(1-\frac{2GM_{NS}}{R_{NS}c^2}\right)}} \frac{E}{11.6}\simeq(2.4-2.9)\times10^{12} {\rm G},$$

\noindent depending on the model used (\emph{gabs} or \emph{cyclabs}) and using $M_{NS}=1.4M_{\sun}$ and $R_{NS}=15$ km \citep{sul2011} for the neutron star radius and mass estimates, respectively. We expect that at the low luminosity of $\sim10^{35}$ erg s$^{-1}$ the accreted matter is falling down to the neutron star surface, where most of the energy is released and the cyclotron absorption line is formed \citep[see, e.g.][]{bas76}. Finally, it is important to note that the cyclotron line in X\,Persei was found to be significantly broader than typically observed in X-ray pulsars \citep{cob02,fil2005}. The unusual broadness may be explained by insufficient knowledge of  spectral continuum in the broad energy band. In particular, approximation of the spectrum with other components can lead to the artificial deficit of photons in the region where the spectral components overlap \citep[see, e.g., Fig.8 in][ or current paper of Doroshenko et al. 2012]{disal98}. This region roughly corresponds to the energy of the cyclotron line that can lead to the distortion of its parameters, particularly the width and depth of the line.

\subsection{Correlation of X-ray and optical light curves}

\begin{figure}
\includegraphics[width=\columnwidth,bb=15 255 515 710,clip]{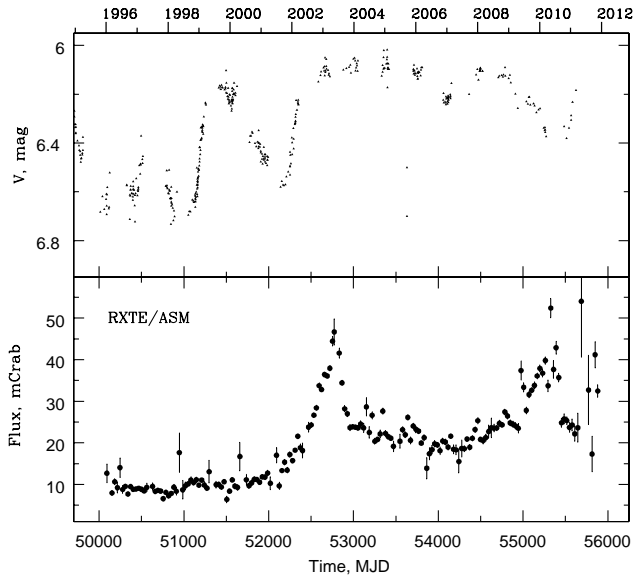}
\caption{Optical light curve of X\,Persei in 1996-2011 in $V$-band (top panel) along with the \textit{ASM/RXTE} light curve (bottom panel). \label{fig:optic_lc}}
\end{figure}

Since its discovery X\,Persei has been regularly observed in different wavebands. Particularly, it was found that long-term fluctuations in the optical brightness ($V\sim6.7-6.1$) are accompanied by the cyclic variability in the peak ratio of H$_{\alpha}$ and He\,I lines with periods varying between $\sim 1-10$ years \citep[see][and references therein]{cla01}. The optical variability of the source is usually interpreted as a change of the mass content and distribution in the circumstellar disk due to the presence of one armed density wave. The lack of correlation between the strength of the line and continuum emission demonstrates that the disk variability involves a redistribution of material within the disk, leading to changes in the radial density gradient \citep{cla01}.

\begin{figure}
\includegraphics[width=0.96\columnwidth,bb=90 280 530 690,clip]{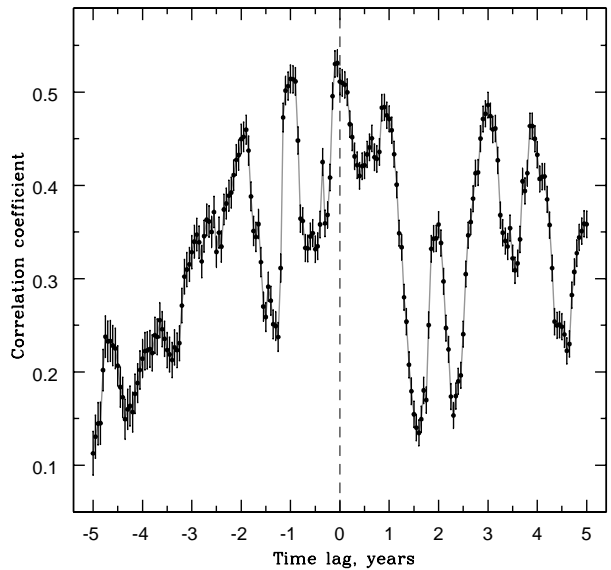}
\caption{Correlation function between optical and X-ray light curves. \label{fig:crosscor}}
\end{figure}

Previous X-ray observations showed a number of cases when a rise of the optical brightness of the source wasn't followed by a corresponding rise of the X-ray flux. This was explained by the truncation of the Be star decretion disk by the tidally-driven eccentric instability \citep[see, e.g.,][]{okazaki01}. Truncation at the 3:1 resonant radius produces a wide gap between the disc outer radius (0.47$a_x$, $a_x$ is an orbital separation) and the periastron distance (0.89 $a_x$) \citep{cla01}. This wide gap depress the accretion even at periastron, naturally leading to a low X-ray luminosity. In addition to this, \citet{cla01} report an episode of complete disc loss during the period 1988 May -- 1989 June. The time needed for the outer radius disk to reach the truncation radius was estimated to be about $\sim 12 (\alpha/0.2)^{-1}$ yr after the disc formation begins \citep[$\alpha$ is the Shakura-Sunyaev viscosity parameter,][]{shakura73}. Thus one can hardly expect any X-ray -- optical correlation before the year 2000.

Regular monitoring of the system by \textit{ASM/RXTE} since 1996 gives us  a unique chance to test this hypothesis and to investigate the X-ray -- optical dependance. Fig.\ref{fig:optic_lc} shows the evolution of the system magnitude in $V$-band (upper panel) along with the X-ray evolution (bottom panel). It is clearly seen that the large variability of the source magnitude has only a weak reflection in the X-ray flux before 2002, when the rise of the optical flux is followed by a major outburst in X-rays. Given the discussion above, this rise in the X-ray flux can be interpreted as the interaction of the neutron star with the fully recovered disk. The maximum rise  of the X-ray flux in 2003 is followed by a rapid decrease to the level still two times higher than the pre-outburst level, while the optical flux stays at a rather high level until a small drop in 2010 and a subsequent rise in 2011. The drop of the X-ray flux in 2004 is probably related to the disruption of the outer layers of the disk, and the second rise of the X-ray flux is in good agreement with the estimated 5.7 years for the disk growth peiord \citep{cla01}.

In order to test qualitatively the  correlation between optical and X-ray light curves with a possible non-zero time lag, we performed an interpolated correlation analysis  using the improvements proposed by  \citet{gas87} and \citet{wht94}. In order to increase the statistic we have binned \textit{ASM/RXTE} data into 10 day bins.
The resulting Interpolated Cross-Correlation Function (ICF) is shown in Fig.\ref{fig:crosscor}. These data show a tentative correlation between X-rays and optical flux with a zero time lag (with a correlation coefficient $R=0.53 \pm 0.01$). The pronounced peaks at time lags of $\pm1, \pm 2, 3, 4$ years are most probably artificial ones due to the internal \textit{ASM/RXTE} one year periodicity  related to the motion of Earth around the Sun \citep{wen06} and also the sampling of the optical observations after 2003 (see Fig.\ref{fig:crosscor}). These facts can also lead to the widening of the ICF, thus despite the quite large correlation coefficient we prefer to call this correlation a tentative one.

\section{Conclusions}

We presented here the comprehensive temporal and spectral analysis of the X-ray pulsar X\,Persei using data of the \textit{RXTE}, \textit{INTEGRAL} and \textit{XMM-Newton} observatories. The main results can be summarized as:

 -- a strong outburst activity was detected from the source in soft and hard X-rays during 2001-2011. During this period the source luminosity twice reached its maximum for the last $\sim30$ years, $L_X\simeq1.2\times10^{35}$ \ergs;

 -- this activity was accompanied by drastic changes of the pulse period: a $\sim35$-years period of the spin-down changed to spin-up with an average rate of $\dot P/P \simeq -3.6\times10^{-4}$ yr$^{-1}$, several times higher than previous rates of spin-up and spin-down; moreover, this acceleration of the rotation of the neutron star is apparently not in steady state, but depends on the source flux; the observed behaviour of the pulse period and its quantitative properties are naturally explained in terms of the current theory of the quasi-spherical accretion in low-luminosity X-ray pulsars \citep{sha12};

 -- at the same time, no significant variability of the spectral parameters with the source flux was found during the rising part of the first outburst in 2001-2003; the source spectrum in the $4-25$ keV energy band can be well described by a combination of a thermal component (with a temperature of $kT\simeq(1.6-1.8)$ keV) and non-thermal component(powerlaw with the photon index of $\Gamma\simeq2$);

 -- using deep observations of the \textit{INTEGRAL} observatory we reconstructed a broadband spectrum of X\,Persei in the $4-200$ keV energy range and, for the first time, significantly detected hard X-ray emission from the source up to $\sim160$ keV;

 -- a significant detection of the cyclotron absorption line in the source spectrum allowed us to estimate the magnetic field strength on the neutron star surface as $B_{NS}\simeq(2.4-2.9)\times10^{12} {\rm G}$ depending on the model used (\emph{gabs} or \emph{cyclabs}); at the same time an insufficient knowledge of the spectrum continuum in the broad energy band can lead to  distortion of the cyclotron line parameters (particularly its width and depth may be affected);

 -- the observed variations of the X-ray flux are in good agreement with predictions of the viscous decretion disk of the Be star \citep{cla01}, a quantitative study of the X-ray -- optical cross-correlation shows only a tentative correlation with a zero time lag.

\section*{Acknowledgments}

Authors thank N.Shakura and K.Postnov for the discussion on results of the paper, and M.Tuerler for the discussion on the details of the ICF method. This work was supported by the program ``Origin, Structure and Evolution of the Objects in the Universe'' by the Russian Academy of Sciences, grants NSh-5603.2012.2 from the President of Russia, grants 11-02-01328 and 11-02-12285-ofi-m-2011 from Russian Foundation for Basic Research, State contract 14.740.11.0611 and the Academy of Finland grant 127512. The research used the data obtained from the HEASARC Online Service provided by the NASA/GSFC, INTEGRAL Science Data Center and Russian INTEGRAL Science Data Center. The results of this work are partially based on observations of the INTEGRAL observatory, an ESA project with the participation of Denmark, France, Germany, Italy, Switzerland, Spain, the Czech Republic, Poland, Russia and the United States. We also acknowledge with thanks the variable star observations from the AAVSO International Database contributed by observers worldwide and used in this research. AL and MC acknowledged the support from ISSI, Bern, where this work was partially done.

\label{lastpage}

\end{document}